\documentclass[aps,prb,twocolumn,showpacs]{revtex4}

\usepackage{graphicx} 

\begin{document}
\topmargin -0.25in
\leftmargin -0.2in
\rightmargin 0.2in

\title{Metallic atomic wires on a patterned dihydrogeneted Si(001)}

\author{Bikash C Gupta}
\author{Inder P Batra}

\affiliation{Department of Physics, University
of Illinois at Chicago, 845 West Taylor Street, Chicago, IL 60607-7059}

\date{\today}

\begin{abstract}

Electronic structure calculations for atomic wire of metals like Al,
Ga and In are performed for a patterned dihydrogeneted Si(001):1
$\times$ 1 in search of structures with metallic behavior. The
dihydrogeneted Si(001) is patterned by depassivating hygrozen atoms
only from one row of Si atoms along the [1$\bar{1}$0] direction.
Various structures of adsorbed metals and their electronic
properties are examined. It is found that Al and Ga atomic wire
structures with metallic property are strongly unstable towards the
formation of buckled metal dimers leading to semiconducting
behavior. Indium atomic wire, however, displays only marginal
preference towards the formation of symmetric dimers staying close
to the metallic limit. The reasons behind the lack of metallic
atomic wires are explored. In addition, a direction is proposed for
the realization of metallic wires on the dihydrogeneted Si(001).
\end{abstract}

 \maketitle

\maketitle

\section{Introduction}

The study of metals on semiconductors dates back to the nineteenth
century and has seen a vigorous recent revival due to tremendous
interest in Nanotechnology. The scanning tunneling microscopy has
enabled us to manipulate atoms, place them at will on different
surface sites to create exotic artificial atomic scale structures
with novel electronic properties.\cite{hoso,crom,shen1,watb} The
placement of metal atoms such as Al, Ga and In on Si(001) may lead
to the formation of low-dimensional structures \cite{shen2},
exhibiting significant new electronic and transport properties.
Atomic scale structures themselves have technological applications
in developing atomic scale devices. \cite{wada} In particular,
realization of a one dimensional metallic nanowire is of great
importance because of its possible use as metallic interconnect in
nano-devices.

There is much current activity in bottoms up approach where free
standing atomic and nanowires for a variety of atoms, {\em e.g.}, K,
Al, Cu, Ni, Au and Si have been studied
\cite{por,por1,tor,tak,hak,sen,sen1,ipb1}. The geometrical
structures of such free standing wires and their electronic
properties have been discussed. A general finding is that a zigzag
structure in the form of an equilateral triangle is most stable
\cite{por,por1,sen,ipb1}. This can be understood as arising
primarily due to the maximization of coordination number for each
atom in a quasi 1D structure. Another structure, a local minimum on
energy surface, but not terribly stable, is a wide angle isosceles
triangle which somehow is reminiscent of the bulk environment. For
example, Si which is four fold coordinated in the bulk (tetrahedral
angle $\sim 109^{\rm o}$) shows \cite{ipb1} a local minimum at an
angle of $\sim 117^{\rm o}$. In general, free standing atomic wires
tend to be metallic (have bands crossing the Fermi level) but these
wires in practice have to be supported. Silicon is the most widely
used substrate for practical applications and the low index
surfaces, Si(001) is the surface of choice. With the downward spiral
toward nano devices, it is desirable to investigate the electronic
properties at the lowest possible coverages. It is in this context
that the study of metals like Al, Ga and In at submonolayer
coverages on Si(001) take on the added importance. The interaction
of metal nanowires with substrate can significantly alter the
electronic properties, and not always in the desired direction.

There continues to be a persistent search for metallic nanowires on
clean Si(001): 2$\times$ 1 surface as well as on the hydrogen
terminated Si(001): 2 $\times$ 1 surface
\cite{watb,ipb2,ipb3,wat1,wat2,him1,him2}. In an early study
\cite{ipb2}, Batra proposed the formation of a zigzag atomic chain
of Al on Si(001): 2$\times$1 but energetically it was not the most
favored structure. In recent studies, it was shown that the zigzag
Al chain is hard to fabricate as it is energetically 1.6 eV higher
than the most favorable structure. However, this chain does not
undergo a Peierls distortion and remains semimetallic in character
\cite{ipb3}.

Recently, the hydrogen terminated Si(001) has become one of the
surfaces of choice for growing atomic scale structures. The hydrogen
terminated Si(001) can have various reconstructed patterns such as
$2\times 1$, $3\times 1$ and $1\times 1$ depending on the hydrogen
coverage and the experimental
environment.\cite{bol1,taut,rag1,bol2,bol3,rag2,mori} Watanabe {\em
et al.} \cite{wat1} explored the growth of Ga on the patterned
monohydride Si(001): 2$\times$ 1. Using STM tip a monohydried
Si(001) may be patterned by removing hydrogen atoms on a chosen row
of surface Si atoms either along [1$\bar{1}$0] or [110] direction.
Watanabe {\em et al.}\cite{wat1,wat2} examined several possible
structures of Ga on such a patterned monohydried Si(001): 2$\times$1
surface and they found one dimensional structures made of small Ga
clusters. However, the structures turned out to be either
semimetallic or semiconducting.

Recently it has been shown in an experiment that an ideal hydrogen
terminated Si(001): $1\times 1$ surface can be achieved by
wet-chemical etching. \cite{mori} Furthermore, it is known that with
the help of a STM tip, some selected hydrogen atoms from the surface
may be desorbed. We therefore, explore the adsorption of Al, Ga and
In on a patterned dihydrogeneted Si(001): 1$\times$1 where Si atoms
on a single row along the [1$\bar{1}$0] direction are depassivated.
In other words, the Si atoms on a single row along the [1$\bar{1}$0]
direction have two dangling bonds each while all other Si atoms on
the surface are saturated with hydrogen atoms. Various possible
structures for Al, Ga and In on the dihydrogeneted Si(001):
1$\times$1 and their properties are studied and a direction is
proposed for the realization of a metallic atomic wire on the
dihydrogeneted Si(001): 1$\times$1.

\begin{figure}[ht]
\includegraphics[scale=0.4]{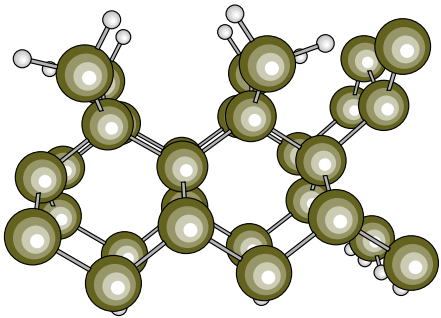}
\caption{Silicon and hydrogen atoms within the supercell
(3$\times$2) for the patterned dihydrogeneted Si(001):1$\times$1.
The large and small circles represent Si and hydrogen atoms
respectively. The top layer Si atoms on the rightmost row have two
dangling bonds each as they are not passivated by hydrogen atoms.}
\label{fig1}
\end{figure}

\begin{figure}[ht]
\includegraphics[scale=0.4]{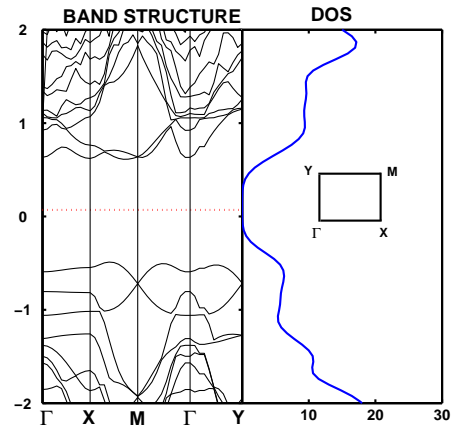}
\caption{Band structure (left panel) and density of states (right
panel) for the patterned dihydrogeneted Si(001) surface shown in
Fig.~\ref{fig1}. The dotted line represents the fermi level. The
$\Gamma$, X, M and Y points are shown in the inset of the right
panel.} \label{fig2}
\end{figure}

\section{Method}

First principle total energy calculations were carried out within
the density functional theory at zero temperature using the VASP
code \cite{kres}. The wave functions are expressed by plane waves
with the cutoff energy $|k + G|^2 \le 250$ eV. The Brillouin Zone
(BZ) integrations are performed by using the Monkhorst-Pack scheme
with 4$\times$4$\times$1 $k$-point meshes for 3$\times$2 primitive
cells. Ions are represented by ultra-soft Vanderbilt type
pseudopotentials and results for fully relaxed atomic structures are
obtained using the generalized gradient approximation (GGA). The
preconditioned conjugate gradient method is used for the wave
function optimization and the conjugate gradient method for ionic
relaxation.

The Si(001) surface is represented by a repeated slab geometry. Each
slab contains five Si atomic planes. The bottom layer Si atoms are
passivated by hydrogen atoms. In addition, within the supercell, two
consecutive rows of Si atoms extending along the [1$\bar{1}$0]
direction on the top layer are passivated with Hydrogen atoms (see
Fig.~\ref{fig1}). Consecutive slabs are separated by a vacuum space
of 9 \AA. The Si atoms on the top four layer of the slab and
hydrogen atoms attached to top layer Si atoms are allowed to relax
while Si atoms in the bottom layer of the slab and the passivating
Hydrogen atoms are kept fixed to simulate the bulk like termination.
The convergence with respect to the energy cutoff and the number of
$k$ points for similar structures has been examined earlier
\cite{ipb3}.

\section{results and discussions}

The lowest energy structure of the patterned dihydrogeneted Si(001)
surface is shown in Fig.~\ref{fig1}. The surface retains its unit
periodicity along the y[1$\bar{1}$0] direction. The unsaturated Si
atoms on the surface (third row of Si atoms on the top layer in
Fig.~\ref{fig1}) forms a dangling bond wire extending along the y
direction. Figure~ \ref{fig2} shows the band structure (left panel)
and the density of states (right panel) for the dangling bond wire.
A wide band gap ($\sim$1.3 eV) around the fermi level indicates that
the surface is semiconducting in nature. The band gap is reflected
in the density of states plot with vanishing density of states
around the fermi level. This is in contrast to the metallic nature
of dangling bond wire on monohydride Si(001): 2$\times$1
surface.\cite{wat1,wat2} The reason behind the non-metallic nature
of the dangling bond wire on the dihydrogeneted Si(001) is that the
unit cell has two free electrons and they are fully accommodated in
a single band below the fermi level.

Metal atoms are adsorbed on the surface shown in Fig.~\ref{fig1} to
examine the possibility of the formation of a metallic nanowire
supported on the Si substrate. Experiments have shown that the Al
and Ga atoms can diffuse easily \cite{shen2,hash} on the hydrogen
terminated surface and therefore, the exposed Al, Ga and In atoms
are expected to diffuse and nucleate around the Si dangling bonds on
the surface. Thus it may be possible to form nanowire of metal atoms
supported on the substrate. We are in search of a supported atomic
wire that will be metallic in character. The adsorption of metals
like Al, Ga and In is studied as a part of this search.


\begin{figure}[ht]
\includegraphics[scale=0.4]{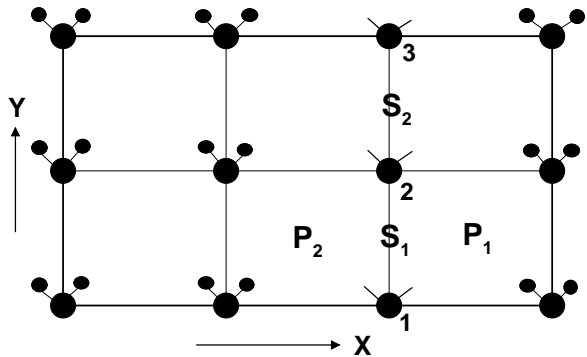}
\caption{Schematic representation of the atomic positions on the top
of the slab within the supercell. Small circles represent the
hydrogen atoms on the top of the slab and large circles represent
the fiest layer Si atoms. The Si atoms with dangling bonds are
marked as 1, 2 and 3 respectively. The sites ${\rm S_1}$ and ${\rm
S_2}$ have identical surrounding and similarly sites ${\rm P_1}$ and
${\rm P_2}$ have identical environment.} \label{fig3}
\end{figure}

\begin{figure}[ht]
\includegraphics[scale=0.4]{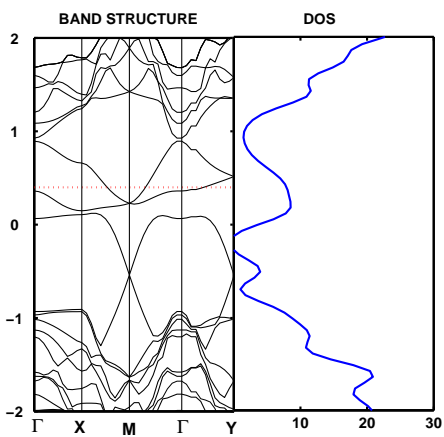}
\caption{The band structure (left panel) and the density of states
(right panel) corresponding to the ${\rm S}_1{\rm S_2}$
configuration of Al. The dotted line shows the fermi level.}
\label{fig4}
\end{figure}

\begin{figure}[ht]
\includegraphics[scale=0.4]{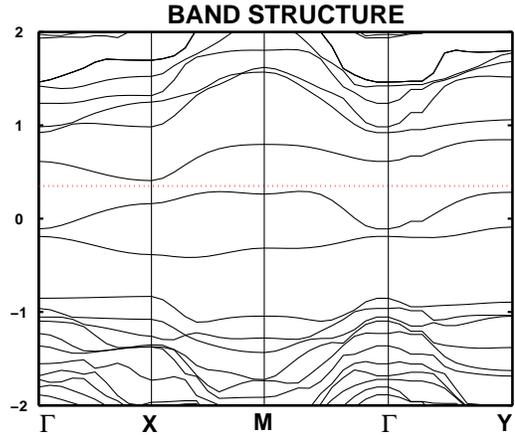}
\caption{The band structure corresponding to the ${\rm P}_1{\rm
P_2}$ configuration of Al. The fermi level is indicated by the
dotted line.}
\label{fig5}
\end{figure}

\begin{figure}[ht]
\includegraphics[scale=0.4]{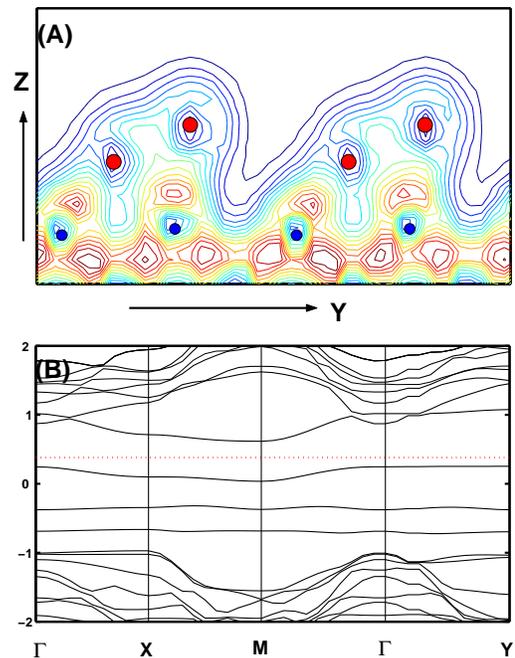}
\caption{This corresponds to the most favorable for the Al-dimer
configuration: (A) The charge density plot on the y-z plane
including the Al atoms and (B) the band structure where the dotted
line indicates the fermi level.} \label{fig6}
\end{figure}

Here we consider the Al adsorption on the patterned dihydrogeneted
Si(001) surface shown in Fig.~\ref{fig1}. The surface offers various
possible sites for the Al adsorption. Two different kinds of sites
are denoted by S (${\rm S}_1$ or ${\rm S}_2$) and P (${\rm P}_1$ or
${\rm P}_2$) respectively as shown in Fig.~\ref{fig3}. The sites
vertically above the hydrogen free Si atoms (Si atoms with dangling
bonds) are denoted as T (${\rm T}_1$ or ${\rm T}_2$). The
configurations considered here are ${\rm T}_1{\rm T_2}$ (one Al is
placed on top Si atom marked as 1 and the other Al atom is placed on
top of the Si atom marked as 2), ${\rm S}_1{\rm S_2}$ (one Al is
placed at ${\rm S_1}$ site and the other one is placed at ${\rm
S_2}$ site) and ${\rm P}_1{\rm P_2}$ (one Al is placed at ${\rm
P_1}$ site and the other is placed at ${\rm P_2}$ site)
respectively. For the ${\rm T}_1{\rm T_2}$ configuration, the Al
atoms are allowed to move along the z direction only, this
configuration is found to be least favorable. The total energies of
other configurations are calculated with respect to the total energy
for the ${\rm T}_1{\rm T_2}$ configuration. The ${\rm S}_1{\rm S_2}$
and ${\rm P}_1{\rm P_2}$ configurations are more favorable than the
${\rm T}_1{\rm T_2}$ by 0.42 and 0.71 eV respectively. The ${\rm
S}_1{\rm S_2}$ configuration is interesting because the band
structure and the density of states (see Fig.~\ref{fig4}) for this
structure indicate the metallic behavior of Al atomic wire extending
along the y direction. On the other hand, the Al structure with the
${\rm P}_1{\rm P_2}$ configuration shows non-metallic behavior (see
a band gap around the fermi level in Fig.~\ref{fig5}). For the ${\rm
S}_1{\rm S_2}$ configuration, each Al atom forms two bonds with two
Si atoms (each Si-Al bond length is $\sim 2.6$ \AA~) and the surface
retains its unit periodicity along the y direction. Therefore, the
single free electron of the Al atom in the unit cell is responsible
for the partially filled bands crossings through the fermi level.
Consequently the density of states increases around the fermi level
and the atomic wire (with ${\rm S}_1{\rm S_2}$ configuration)
becomes metallic in character. For the ${\rm P}_1{\rm P_2}$
configuration, we note that Al atoms make strong bonds with Si atoms
(bond length $\sim 2.5$ \AA~ while they fail to make a bond among
themselves (distance between two Al atoms $\sim 3.2$ \AA). In this
configuration, the periodicity along y direction is doubled compared
to that for the ${\rm S}_1{\rm S_2}$ configuration. Two electrons
from two Al atoms become non-itinerant in the unit cell, prefer to
be occupied by a single band and hence the Al structure (with ${\rm
P}_1{\rm P_2}$ configuration) extending along the y direction
becomes non-metallic. Similar to the zigzag Al chain considered
earlier \cite{ipb2,ipb3} on the bare Si(001): 2$\times$1, we also
considered a zigzag Al chain configuration by displacing the Al atom
at ${\rm S_1}$ site by $\Delta$ along the +ve x direction and the Al
atom at ${\rm S_2}$ site by $\Delta$ along the -ve x direction. This
zigzag chain configuration turned out to be unstable and readily
reverted to the ${\rm S}_1{\rm S_2}$ configuration.

\begin{figure}[ht]
\includegraphics[scale=0.4]{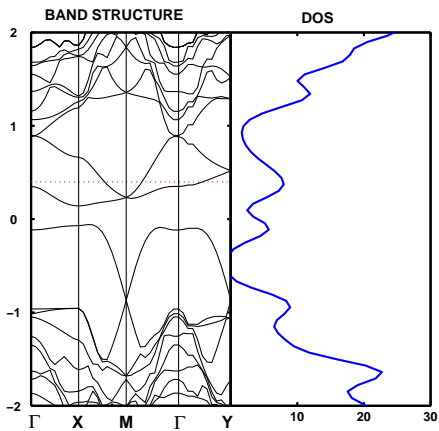}
\caption{The band structure (left panel) and the density of states
(right panel) corresponding to the ${\rm S}_1{\rm S_2}$
configuration of Ga. The fermi level is shown by the dotted line.}
\label{fig7}
\end{figure}

\begin{figure}[ht]
\includegraphics[scale=0.4]{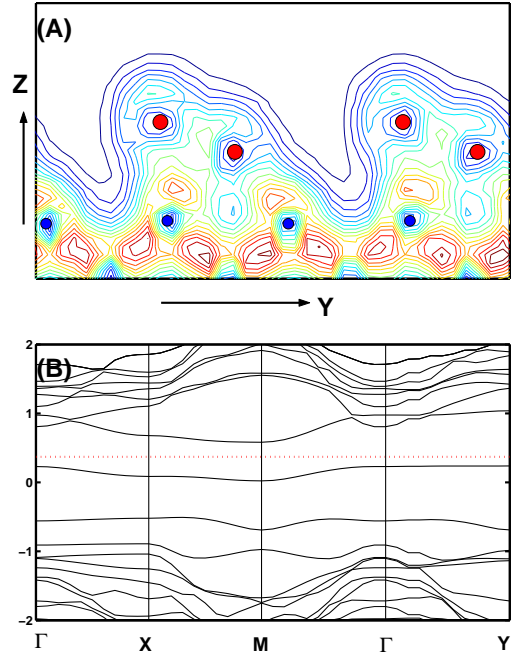}
\caption{This corresponds to the most favorable configuration of Ga
(Ga-dimer configuration): (A) The charge density plot on the y-z
plane including the Ga atoms and (B) the band structure where the
fermi level is indicated by the dotted line.}
\label{fig8}
\end{figure}

However, the most favorable configuration is the one where the Al
atom at ${\rm S_1}$ site is shifted along the positive y direction
by 0.2 \AA~ while  that at the ${\rm S_2}$ is shifted along the
negative y direction by 1.6 \AA~ to form a buckled Al dimer with a
bond length of $\sim 2.6$ \AA.~ The buckled Al dimers can be seen in
charge density plot in the Fig.~\ref{fig6}(A) where the large
circles represent Al atoms while the small circles represent Si
atoms (just below the Al wire). We will call this configuration as
Al-dimer configuration and has a behavior akin to buckled Si dimer
on Si (001). This is clear from the charge density plot in
Fig.~\ref{fig6}(A). The total energy of this configuration is -0.94
eV compared to the ${\rm T}_1{\rm T_2}$ configuration, {\em i.e.},
this Al-dimer configuration is favorable over ${\rm S}_1{\rm S_2}$
configuration by 0.52 eV. This energy is gained mostly due to the
optimization of Al-Si bonds and the superiority of Al-Al bond with
respect to Al-Si bond. The nature of the Al wire extending along the
y direction becomes non-metallic as can be seen from the band
structure with an energy gap around the fermi level (see Fig.~
\ref{fig6}(B)). We therefore, conclude that the metallic Al nanowire
( corresponding to the configuration ${\rm S}_1{\rm S_2}$ on the
dihydrogeneted Si(001) can not be achieved in practice.

We nest consider the Ga adsorption on the same patterned dihydride
Si(001). It is known that at some coverages, Ga behaves in a
different way compared to Al \cite{ipb3}.  The ${\rm T}_1{\rm T_2}$
configuration is the least favorable one. However, unlike the case
of Al, the ${\rm S}_1{\rm S_2}$ configuration for Ga (total energy
-0.41 eV compared to  ${\rm T}_1{\rm T_2}$ configuration) is
slightly more favorable by 0.02 eV compared to the ${\rm P}_1{\rm
P_2}$ configuration. The Ga atomic wire with the  ${\rm S}_1{\rm
S_2}$ configuration is metallic with a peak for density of states
around the fermi level (see Fig.~\ref{fig7}). However, the most
stable configuration is very similar to that we found for Al. Two Ga
atoms form buckled dimer as can be seen from the charge density plot
in Fig.~\ref{fig8}(A) (the large circles represent the Ga atoms and
the small circles represent Si atoms just below the Ga atoms). This
most favorable configuration is denoted as Ga-dimer configuration.
The atomic wire made of buckled Ga dimers extending along the y
direction is non-metallic in nature and this can be seen from the
band structure plot in Fig.~\ref{fig8}(B). Similar to the Al case,
we therefore, conclude that stable metallic Ga wire can not be
realized on patterned dihydrogenetd Si(001). We however, notice that
the total energy difference between the Ga-dimer configuration and
the ${\rm S}_1{\rm S_2}$ configuration is 0.41 eV which is lower by
0.1 eV compared to that for the Al case. The reduction in the total
energy difference between the ${\rm S}_1{\rm S_2}$ and dimer
configuration for Ga encourages us to study the In adsorption on the
same patterned dihydrogented Si(001).

\begin{figure}[ht]
\includegraphics[scale=0.4]{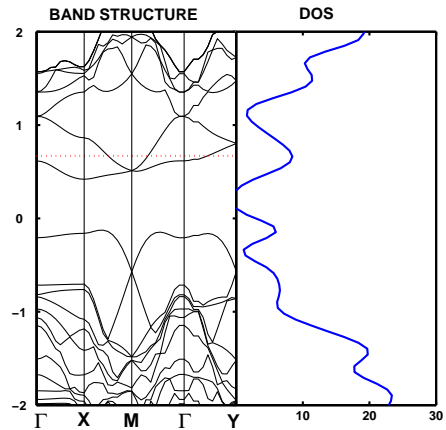}
\caption{The band structure (left panel) and the density of states
(right panel) corresponding to the ${\rm S}_1{\rm S_2}$
configuration of In. The fermi level is represented by the dotted
line.}
\label{fig9}
\end{figure}

\begin{figure}[ht]
\includegraphics[scale=0.4]{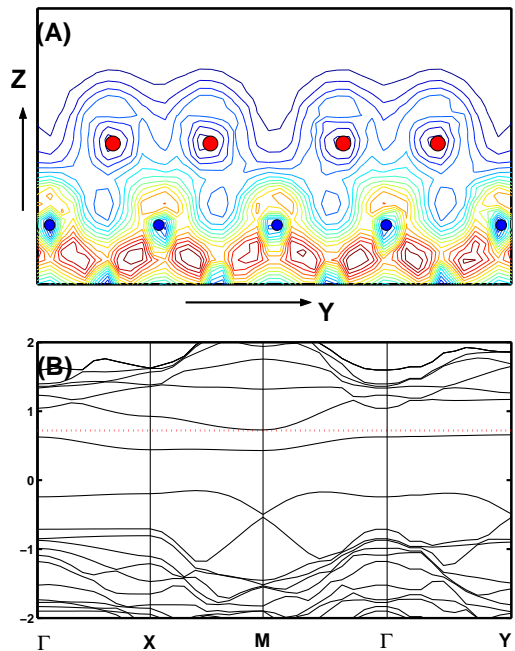}
\caption{This corresponds to the most favorable configuration for
In: (A) The charge density plot on the y-z plane including the In
atoms and (B)The band structure, dotted line represents the fermi
level. } \label{fig10}
\end{figure}

Unlike the case of Al and Ga the ${\rm P}_1{\rm P_2}$ is the least
favorable configuration for In. The total energy for this
configuration compared to the ${\rm T}_1{\rm T_2}$ configuration is
+0.08 eV. The total energy for the ${\rm S}_1{\rm S_2}$
configuration compared to the ${\rm T}_1{\rm T_2}$ configuration is
-0.20 eV. The In atomic wire with the ${\rm S}_1{\rm S_2}$
configuration is metallic in character with large density of states
around the fermi level and this can be seen from the band structure
(left panel) and density of states (right panel) plot in
Fig.~\ref{fig9}. The most favorable configuration for In is a
non-buckled In dimer as seen from the charge density plot in
Fig.~\ref{fig10}(A) where the In atoms (large circles in the figure)
at ${\rm S_1}$ and ${\rm S_2}$ positions move towards each other by
0.3 \AA~ to form a weak dimer. Consequently, the nanowire consisting
of unbuckled weak In dimers becomes non-metallic (see the band
structure in the Fig.~\ref{fig10} (B)). We note that the total
energy for this configuration is -0.27 eV which is more favorable to
the ${\rm S}_1{\rm S_2}$ configuration only by 0.07 eV. At room
temperature the thermal energy is expected to be sufficient to break
these weak dimer bonds leading to metallic behavior of In wires on
the dihydrogenetd Si(001). Here we add that there are situations
where the experimentally observed structure corresponds to some
local minimum energy structure. \cite{ipb3,bour} Therefore, an
experiment on this system is desirable to confirn if In atomic wire
can be realized.

\section{conclusion}
First principle electronic structure calculations are performed to
examine the possibility for the formation of stable metallic atomic
wires on the dihydrogenetd Si(001). Adsorption of metals like Al, Ga
and In are considered for this purpose. We found that the Al and Ga
nanowire configurations with metallic character are strongly
unstable towards the formation of buckled metal dimers leading to
semiconducting behavior. However, the metallic In wire is weakly
unstable because the total energy corresponding to the metallic wire
configuration is very close to the most favorable non-metallic
(weakly dimerised) configuration. The thermal energy may be able to
break the weak bonds between In atoms and thus a metallic In wire
may be realized on Si(001). Our results clearly indicates that as we
go from Al to In via Ga, the metallic nanowire configuration
approaches towards the most favorable one. We are hopeful that this
work will encourage further experimental studies of In atomic wires
on a patterned dihydrogeneted Si(001).

\end{document}